\documentclass[preprint,floats,aps,epsfig,nofootinbib,amssymb]{revtex4}

\usepackage{slashed}
\usepackage{slashed}
\usepackage{graphicx,color}
\usepackage{epsfig}
\usepackage{subfigure}
\usepackage{epsfig}
\usepackage{dcolumn}
\usepackage{bm}
\usepackage{color}

\def\lsim{\mathrel{\rlap{\lower4pt\hbox{\hskip1pt$\sim$}}
    \raise1pt\hbox{$<$}}}         
\def\gsim{\mathrel{\rlap{\lower4pt\hbox{\hskip1pt$\sim$}}
    \raise1pt\hbox{$>$}}}         

\newcommand{\be}{\begin{equation}}
\newcommand{\ee}{\end{equation}}
\newcommand{\bea}{\begin{eqnarray}}
\newcommand{\eea}{\end{eqnarray}}

\begin{document}

\vspace*{-5.8ex}
\hspace*{\fill}{NPAC-12-17}

\vspace*{+3.8ex}

\title{Parity- and Time-Reversal Tests in Nuclear Physics}

\author{David Hertzog$^{1}$}
\email{hertzog@uw.edu}
\author{Michael J. Ramsey-Musolf$^{2,3}$ }
\email{mjrm@physics.wisc.edu}
 \affiliation{$^1$ Department of Physics, University of Washington, Seattle, WA 98195 USA\\
 $^2$ Department of Physics, University of Wisconsin-Madison, Madison, WI 53706, USA\\
$^3$California Institute of Technology, Pasadena, CA USA\\
 }

\vspace{3cm}

\begin{abstract}
Nuclear physics tests of parity- and time-reversal invariance have both shaped the development of the Standard Model and provided key tests of
its predictions. These studies now provide vital input in the search for physics beyond the Standard Model. We give a brief review of a few key experimental and theoretical developments in the history of this sub-field of nuclear physics as well as a short outlook, focusing on weak decays, parity-violation in electron scattering, and searches for permanent electric dipole moments of the neutron and neutral atoms.

\end{abstract}

\maketitle
\section{Introduction}
\label{sec:intro}

Studies of fundamental symmetries in nuclei, hadrons, and atoms have played a vital role in the development and testing the Standard Model (SM) of electroweak and strong interactions. The observation of parity-violation (PV) in the decay of polarized $^{60}$Co nuclei, together with the analogous observation of PV in the decay of the pion, provided the experimental foundation for the $(V-A)\times (V-A)$ structure of the SM charged current (CC) interaction. During the same era, the search for a permanent electric dipole moment (EDM) of the neutron as first proposed by Ramsey and Purcell in 1950 launched a half century of EDM searches that have now placed stringent limits on possible parity (P) and time reversal (T) violation in the strong interaction. Two decades later, the measurement of the PV asymmetry in deep inelastic scattering of polarized electrons from deuterium singled out the SM theory of neutral weak interactions from competing alternatives well before the discovery of the weak gauge bosons at CERN. In more recent times, the approximate chiral symmetry of Quantum Chromodynamics for light quarks has lead to a number of predictions that have been confirmed with increasing precision in hadronic and few-body systems.

Today, SM predictions have been confirmed by a plethora of electroweak and strong interaction precision tests, carried out at energies ranging from atomic scales to those of high energy colliders. The quest is now to determine the larger framework that contains the SM (physics beyond the Standard Model, or BSM) and to explain how the nonperturbative dynamics of the strong interaction give rise to the observed properties of hadrons and nuclei. In both cases, the history of fundamental symmetry tests in nuclear physics, coupled with the substantial advances in both theory and experiment, point to a key ongoing role for these studies in uncovering and elucidating the basic laws of nature.

In what follows, we give a brief overview of the history of fundamental symmetry tests in nuclear physics, focusing on P and T. We then survey the outlook for the coming decade, highlighting the important questions these studies may address and their complementarity with BSM physics searches at the Large Hadron Collider. In section \ref{sec:pvcc} we focus on parity-violation in charged current interactions (primarily weak decays) while in section \ref{sec:pvnc} we give the corresponding discussion for neutral current studies. Section \ref{sec:trv} contains a history of time-reversal invariance tests, emphasizing largely the EDM efforts. In Section \ref{sec:out} we provide an outlook for the future. The flavor of this article is largely qualitative and retrospective, keeping technical details to a minimum. For reviews having more of a research emphasis, see, {\em e.g.}   \cite{Musolf:1993tb,Beck:2001dz, Erler:2004cx,Beise:2004py,RamseyMusolf:2006dz,RamseyMusolf:2006vr}     .

\section{Parity-Violation in Charged Current Interactions}
\label{sec:pvcc}
The discovery of PV in the $\beta$-decay of polarized $^{60}$Co by Wu {\em et al.}\cite{Wu:1957my} represented a milestone in the development of the SM and in nuclear physics. This observation, and the nearly concurrent observation of PV in pion decay by Garwin, Lederman, and Weinrich\cite{Garwin:1957hc}, followed an earlier proposal by Lee and Yang that the weak interaction would violate parity\cite{Lee:1956qn}. The presence of PV would allow for a resolution of the \lq\lq $\tau$-$\theta$" puzzle, associated with the observation of two- and three-pion final states in the decay of two strange particles---the $\tau^+$ and $\theta^+$, respectively---having nearly the same masses and lifetimes. Since the $2\pi$ and $3\pi$ states have opposite parity, this situation could only be resolved if either the two parent strange mesons were distinct particles or if the various pionic final states arose from the decay of a single particle in the presence of PV.

Lee and Yang discussed a number of observables that could conclusively demonstrate parity violation, including a non-vanishing neutron EDM and a directional asymmetry $A_\beta$ in the $\beta$-decay of polarized  nuclei:
\be
\label{eq:beta1}
A_\beta = \frac{\int_0^{\pi/2}\ d\theta I(\theta)-\int_{\pi/2}^\pi\ d\theta I(\theta)}{\int_0^{\pi/2}\ d\theta I(\theta)+\int_{\pi/2}^\pi\ d\theta I(\theta)}\ \ \ ,
\ee
where $I(\theta)d\theta$ is the number of $\beta$ particles emitted in an infinitesimal range $d\theta$ around a direction $\theta$ with respect to the nuclear polarization axis. In addition to discussing this asymmetry, Lee and Yang also suggested the measurement of a directional asymmetry in the decay of muons emitted in pion decay. Parity violation in the latter process would imply polarization of the muon in the direction of its momentum, thereby defining a polarization axis analogous to that of the polarized nuclei. The angular distribution of the emitted electron (or positron) would then be asymmetric with respect to the direction of momentum, leading to an asymmetry similar to that of Eq.~(\ref{eq:beta1}).

Interestingly, a means of searching for the EDM had been proposed by Ramsey and Purcell in 1950\cite{Purcell:1950zz}. Ramsey, Purcell, and Smith subsequently reported the  first limits from such a search at prior to the appearance of Lee and Yang's paper (see Sec. \ref{sec:trv} below), whose publication in turn preceded publication of the EDM result in Physical Review\cite{Smith:1957ht}. The discussion of $\beta$- and pion-decays motivated two groups to search for these effects.
%
The results were reported in back-to-back articles in Physical Review\cite{Wu:1957my,Garwin:1957hc}. Lee and Yang were awarded the 1957 Nobel Prize in physics for \lq\lq their penetrating investigation of the so-called parity-laws..." \cite{LNNobel}. Importantly for the future development of the field, the Appendix of  Ref.~\cite{Lee:1956qn} contained the first expression for the most general $\beta$-decay Hamiltonian that allows for PV as well as the formula for the angular distribution of $\beta$ particles. These expressions were utilized by a plethora of subsequent experimental and theoretical studies, including the seminal paper on time-reveral tests in $\beta$ decay by Jackson, Treiman, and Wyld that appeared the following year\cite{Jackson:1957zz}.

By the time of Glashow's 1961 paper on electroweak unification and gauge symmetry\cite{Glashow:1961tr}, the violation of parity in weak interactions was part of the common body of elementary particle physics knowledge and lead Glashow to exclude certain classes of electroweak symmetries. Weinberg's \lq\lq Model of Leptons" that incorporated the Higgs idea of spontaneous symmetry-breaking\cite{Weinberg:1967tq} (see also the work by Salam\cite{Salam:1968rm}) assumed the purely left-handed nature of charged current (CC) weak interactions and did not even discuss the rationale for putting the right-handed charged leptons into a singlet representation of the non-Abelian gauge group. A discussion of tests of the weak neutral current (NC) will appear in section Ref.~\ref{sec:pvnc}. It is important to emphasize, however, that these tests exploited the parity-violating nature of the weak interaction to filter out the effect of the weak NC interaction from the much stronger electromagnetic interaction.

Given the fundamentally important nature of the first observations of PV in $\beta$-decay and $\pi$-decay, it is worth devoting some space to a brief discussion of the experiments. Lee and Yang's paper triggered a local (Columbia University) response~\cite{Wroblewski:2008zz} by the experimentalists.   Wu suggested a ``simple'' $\beta$-decay experiment.  It would compare the rate of the dominant 310~keV electrons in $^{60}$Co decay with respect to the orientation of the spin of the nucleus.  The electron detection required a scintillation crystal viewed by a photomultiplier tube (PMT), which was a straightforward technique.  However, polarizing the nucleus was beyond local expertise. Fortunately, a method to align nuclear spins in the absence of extremely high external magnetic fields had been proposed by both Gorter and Rose~\cite{GorterRose} and had been demonstrated using different methods by Bleaney et al.~\cite{Bleaney:1954} and Ambler et al.~\cite{Ambler:1953}.  Both used $^{60}$Co decay because its two decay gamma rays from the de-excitation of the excited $^{60}$Ni daughter state were known to have a (non-parity-violating) spatial emission pattern with respect to the nuclear spin axis. Ambler and others from the National Bureau of Standards teamed with Wu to develop an experiment that could measure both the gamma and beta directions as a function of a controllable cobalt nuclear spin state.  A thin layer of $^{60}$Co was deposited on a cerium magnesium nitrate paramagnetic crystal, which was used for cooling to 0.003~K.  At this temperature the thermal energy is below that might required to flip the nuclear spins.  A relatively low-field solenoid surrounding the sample was applied, which consequently aligned the $^{60}$Co nuclei along the field direction; that is, the direction of the spin, not just the alignment, could be set.  Inside the cryostat and just above the crystal, a scintillator was placed to measure the $\beta$-decays, the light from which was transported along a guide to a PMT located outside.  The rate of electrons was counted versus the nuclear spin direction and also from the unpolarized state. The results provided a clear indication of the violation parity in a weak decay.

Well known is the report that Lederman learned of Wu's pre-publication results at a faculty gathering and, following discussions with Wu and Lee, rapidly organized a test together with Garwin and Weinrich at the Nevis cyclotron using a beam of pions and muons.  Pion-to-muon decay in-flight was suggested to emit a  muon with a polarization along the pion momentum axis.  The decay electron from a polarized muon should in turn be aligned along the axis of the muon spin; that is, also along the beam direction.  We ignore the neutrinos here and avoid assigning any bias to the sign of the proposed asymmetry.  The Columbia experiment used muons brought to rest in a carbon target, an assumption being made that the muons would retain their spin orientation during the braking process.  The muon beam was obtained from in-flight pion decays; however, the beam also contained pions that had not yet decayed;  they were removed by a degrader. A coincidence between scintillator counters on both sides of this absorber defined a muon stop. The decay electrons were then counted using a scintillator-absorber-scintillator telescope that viewed the target from behind a shielding wall.  It was gated to count in a $1.25~\mu$s time window starting $0.75~\mu$s after a muon stop.   This generic setup could be used to determine the positive and negative muon lifetimes in various materials, thereby providing information on the Fermi constant and on muon capture rates.  It was not difficult to modify it to test parity by adding a magnetic field surrounding the stopping target with a field orientation transverse to the beam axis.  With appropriate field strength (up to 50~ G),  a muon spin could be made to precess by an appreciable amount prior to and during the observation window.  The counting rate versus muon spin direction followed a $1 + \alpha\cos\theta$ behavior with $\alpha \approx -1/3$ determined as an upper limit.  The data were fit to find the precession frequency and thereby obtain also a value for the gyromagnetic ratio of the muon, giving $g = 2.00 \pm 0.10$.

These experiments were remarkably clean arriving at indisputable findings.  In both cases, the teams incorporated numerous systematic checks and made null measurements with unpolarized samples to test for possible biases in their counting procedures.  They were at times, also lucky. It would not have been possible to carry out the $\beta$-decay experiment without the years of work that had been devoted to polarizing nuclei with a motivation unrelated to parity tests.  In the muon experiment, we now know that polarized muons do retain their incident polarization as they come to rest in some materials, but not fully in others.  The physics relies on material effects, local fields and the formation and destruction of the muonium (a $\mu^+$e$^-$) atom along the way (which had not yet been discovered).   The authors were not unaware of that the muon spin precession characteristics might be influenced by internal as well as external fields.  Indeed the technique of $\mu$SR---muon spin rotation, resonance, or relaxation---is by now a mainstream tool in condensed matter physics.  It was subtly suggested in the concluding remarks to Ref.~\cite{Garwin:1957hc}: ``It seems possible that polarized positive and negative muons will become a powerful tool for exploring magnetic fields in nuclei ... atoms, and interatomic regions.''

Finally, we remark that modern muon experiments have measured the muon lifetime to a precision of 1.0~ppm, obtaining the Fermi constant to 0.5~ppm~\cite{Webber:2010zf}.  Separately, the $g$-factor of the muon has been measured to nearly 0.3~ppb~\cite{Bennett:2006fi} and the value, when compared to the Standard Model, gives a tantalizing 3.6~$\sigma$ discrepancy and a possible hint of BSM physics.  Ironically, the role of parity-violation in the lifetime measurement was a nuisance; it had to be suppressed as much as possible by experimental symmetry and target depolarization techniques to avoid counting bias ``versus space.''  In the $g$-factor effort, it is parity violation that enables the entire method, first by providing a naturally polarized source of muons from pion decay and next by allowing the PV decay to be used as a spin analyzer.  Indeed, the magnetic moment measurement concept was outlined in Lee and Yang's original work.

Apart from theoretical developments, the study of PV in the weak decays of nuclei, hadrons such as the neutron, and charged leptons has become something of an industry in nuclear physics. A full discussion of this history goes beyond the scope of this brief review, so we provide only two recent examples, both of which are being used not only to test the SM but also to probe for indications of possible physics beyond it. First, we consider the decays of polarized muons. The distribution of daughter electrons (positrons) in the decay of the $\mu^-$ ($\mu^+$) as a function of their energy $E_e$ can be characterized by the so-called \lq\lq Michel parameters" \cite{Michel:1949qe,Bouchiat:1957zz,Kinoshita:1957zza}. Of particular interest to the discussion of PV is the spatially anisotropic term in the distribution
\be
\label{eq:michel}
d\Gamma\vert_\mathrm{PV} \sim \mathcal{P}_\mu\xi\cos\theta \left[ (1-x) +\frac{2}{3}\delta (4x-3)\right]
\ee
where $\theta$ again denotes the angle between the direction of muon polarization, with $\mathcal{P}_\mu$ denoting the degree of polarization and $x$ giving the ratio of $E_e$ to its maximum value $\approx m_\mu/2$. Recently, the TWIST collaboration has completed a comprehensive experimental study of the muon-decay distribution, yielding a result for the PV directional asymmetry parameters that agree with SM expectations with $\sim 0.1\%$ precision\cite{Bueno:2011fq,Bayes:2011zza}.
Together with the new results for the parity-conserving component of the energy distribution, the TWIST PV results constrains the possible right-handed muon coupling to be smaller than a part per thousand.

The study of PV in the decays of nuclei and the neutron is providing an equally interesting probe of possible extensions of the SM (for a recent review, see
\cite{Severijns:2006dr}) as well as the determination of parameters that characterize the SM weak interaction. One such parameter is the nucleon axial vector coupling that enters the matrix element of the  axial vector current:
\be
\label{eq:gA}
\langle N| A_\mu (0) | N\rangle = g_A {\bar N} \gamma_\mu\gamma_5\tau_3 N +\cdots
\ee
The coupling $g_A$ can be determined by measuring the PV directional asymmetry in the decay of polarized neutrons
\be
\Gamma\sim A {\vec S}\cdot {\vec p}_\beta +\cdots
\ee
where the asymmetry parameter $A$ is given by
\be
\label{eq:bigAtoLambda}
A=-2\frac{\lambda(1+\lambda)}{1+3\lambda^2}
\ee
where $\lambda= g_A/g_V $ with the vector coupling $g_V$ given by the analog of Eq.~(\ref{eq:gA}) for the vector current. Knowledge of $g_A$ is required for understanding of weak interactions in stars as well as for a determination of the Cabibbo-Kobayashi-Maskawa matrix element $V_{ud}$ when combined with the value of the neutron lifetime\footnote{The most precise value of $V_{ud}$ is obtained from the study of super-allowed Fermi nuclear $\beta$-decays.}.

Recent measurements by the Perkeo II~\cite{Mund:2012fq} and UCNA~\cite{:2012mea} collaborations that exploit cold and ultra-cold neutrons, respectively, have yielded values of $\lambda$ with roughly 0.1\% precision. The design concept for these experiments is worth describing.  Both measure the correlation between the electron momentum and the neutron spin in the decay $n \rightarrow p + e^- + {\bar{\nu}}_e$. The energy-dependent electron emission probability at angle $\theta$ with respect to the neutron polarization is
\be
 W(\theta) = 1 + \frac{v}{c}PA\cos\theta,
\ee
 where $v$ is the electron velocity, $P$ the polarization magnitude, and $A$ the measurable asymmetry. Apart from $\approx1~\%$ corrections, it is related to $\lambda$ as given in Eq.~\ref{eq:bigAtoLambda}.

Neutrons from a reactor are made ``cold'' by scattering off (light) nuclei in a cryogenic moderator.  At the Institute Laue-Langevin in France where the Perkeo~II experiment is performed and a follow-up Perkeo~III is underway, typical neutron kinetic energies are $\sim25$~meV.  They can be polarized to nearly $100\%$ transverse to the beam direction using so-called super-mirror coated bender polarizers in crossed geometry.  Fast spin flippers allow the orientation to be selected at will.  The neutrons pass into the experimental fiducial volume, which is surrounded by a transverse superconducting magnetic field.  Approximately one in $10^5$ decay in a volume that can catch the emitted electron and direct it either along or opposite the neutron spin orientation to a detector that measures its energy.  The asymmetry $A_\beta(E)$ is obtained from large data samples and using appropriate sets of field and spin reversals to remove any bias.

In contrast, the UCNA experiment relies on an ``ultra-cold'' neutron (UCN) source, which is produced at the LANSCE facility at Los Alamos National Laboratory.  A UCN has a kinetic energy below 335~neV ($T~<~4$~mK), which, in more familiar terms, corresponds to neutron speeds below $8~m/s$.  UCNs can be trapped in bottles by gravity and, with their wavelengths exceeding $500~\mathrm{\AA}$, they can readily be guided along certain solid surfaces without significant absorption.  The UCNA experiment directs neutrons along guides from a pulsed source, through a magnetic polarizer, and through an adiabatic fast spin flipper, arriving and being trapped inside a tubular decay volume that is oriented transverse to the spin direction.  A highly uniform magnetic field surrounds this volume and directs decay electrons left or right to detectors that measure both energy and position of hits.

In both cases, the design of the experiment is clearly able to detect a ``left-right'' difference in the decay, but we have omitted the many detailed and beautiful features and ignored the systematic tests that are required to validate the results.  Both experiments have, by now,  determined the asymmetry $A$ to sub-percent precision and together they establish the most precise determination of $g_A$.

\section{Parity-Violation in Neutral Current Interactions}
\label{sec:pvnc}
The modern era of searches for PV in processes involving weak neutral  currents (WNC) has, perhaps, its origins in atomic physics and neutrino reactions. In their  1974 paper \lq\lq Weak Neutral Currents in Atomic Physics"\cite{Bouchiat:1974kt}, Claude and Marianne Bouchiat summarized ideas involving atomic PV that had been in the air for sometime,  dating back to Zeldovich\cite{zeldovich61}, who noted the effect would lead to opposite parity admixtures in atomic states,  and later discussed by F. Curtis-Michel\cite{Michel:1965zz}. The latter paper applied this idea to atomic hydrogen. In the much later work of Ref.~\cite{Bouchiat:1974kt}, the Bouchiats noted that the effect of parity-mixing would be enhanced in heavy atoms by a factor of $Z^2$ due to the effect of the large charge $\propto Z$ on the atomic wavefunction. Moreover, the part of the electron-nucleus WNC interaction that is independent of the nuclear spin would be further enhanced by nuclear coherence, since this interaction is dominated by the coupling to the time-component of a nuclear vector current. Thus, one should expect the signal to grow roughly as $Z^3$. The $Z^2$ factor would be independent of the detailed nature of WNC, while the precise form of the coupling of the $Z^0$ to the nucleus -- the so-called \lq\lq weak charge" or $Q_W$ -- would determine the other factors containing the final power of $Z$.

The latter observation motivated the search for PV effects in heavy nuclei such as Cesium, Bismuth\cite{Baird:1977nm,Lewis:1977nn,Barkov:1978fb,Hollister:1981zz}, Thallium\cite{Wolfenden:1991qq}, and Lead\cite{Emmons:1984vn}. At the same time an experiment was undertaken at SLAC to search for the WNC interaction in high-energy scattering of polarized electrons from deuteron scattering. The quantity of interest is the parity-violating asymmetry
\be
A_{PV} = \frac{N_+ - N_{-}}{N_+ + N_{-}} = \frac{G_F Q^2}{4\sqrt{2}\pi\alpha} F(Q^2, y)
\ee
where $N_+$ ($N_{-}$) is the number of detected electrons with a beam with initially positive (negative) helicity; $Q^2=-q^\mu q_\mu$ with $q$ being the four-momentum transfer; $y$ is the dimensionless energy transfer in the target rest frame; and $G_F$ is the Fermi constant. At the time of the SLAC experiment, there existed a variety of competing models for the WNC interaction, each of which predicted a different dependence on $y$. At the kinematics of the SLAC experiment, the magnitude of $A_{PV}$ is $\sim 10^{-4}$.

The results for $A_{PV}$ appeared in 1978\cite{Prescott:1978tm} and indicated only a mild $y$-dependence, a result consistent with the Weinberg-Salaam model that ruled out competing alternatives. In what then became the SM, the $y$-dependent term in $F(Q^2,y)_\mathrm{SM}$ is proportional to
the vector coupling of the $Z^0$ to the electron,
\be
g_V^e = -1+4\sin^2\theta_W
\ee
with $\theta_W$ denoting the weak mixing angle. Subsequent results were reported the following year. From a fit to the $y$-dependence of $A_{PV}$ the experiment yielded $\sin^2\theta_W=0.224\pm0.020$.

Despite the earlier start of the atomic PV experiments, the first non-zero results for an atomic PV observable were not reported until 1982 in an experiment using Cesium\cite{Bouchiat:1982um}. The subsequent decade witnessed a number of efforts to measure atomic PV observables (see NNN for a  review), as well as new measurements of PV electron scattering asymmetries in quasielastic scattering from $^9$Be\cite{Heil:1989dz} and elastic scattering from $^{12}$C\cite{Souder:1990ia}. In all cases, the goal of the experiments was to test the SM prediction for the WNC and to determine the fundamental couplings. The most precise determination of a PV WNC observable was ultimately reported for a determination of the Cesium weak charge after a decade long effort using an atomic beam\cite{Wood:1997zq}. The experimental error in the PV amplitude extracted from the measured atomic transitions was less than 0.5\% but the final error on $Q_W(\mathrm{Cs})$ was larger due to uncertainties in the atomic theory. Nevertheless, the measurement had a significant impact in particle physics, placing severe constraints on the so-called \lq\lq S-parameter" that were in conflict with the standard versions of technicolor. At the same time, the atomic theorists were challenged to refine their computations, leading to a decade long effort that ultimately yielded a theoretical error bar commensurate with that of experiment\cite{Porsev:2009pr}. The current value for $Q_W(\mathrm{Cs})$ is in spectacular agreement with the SM prediction at the 0.5\% level.

During the same decade preceding the report of the Cesium result, a new program of PV electron scattering (PVES)  experiments was initiated whose focus was on using the by-then known structure of the WNC to probe novel aspects of nucleon substructure. This effort was motivated in part by the \lq\lq spin-crisis" resulting from measurements of nucleon structure functions with polarized leptons. Measurements of the structure function $g_1^p$ could be interpreted in terms of the total contribution of the light-quarks to the spin of the nucleon, $\Delta\Sigma \times \hbar/2$. The results obtained by the EMC collaboration\cite{Ashman:1987hv} implied a magnitude for $\Delta\Sigma\sim 0.3$, in dramatic conflict with the na\" ive quark model picture of the nucleon. In addition, the first moment of $g_1^p$ implied a violation of the Ellis-Jaffe sum rule\cite{Ellis:1973kp} and a non-vanishing value for the strange quark contribution $\Delta s\sim -0.1$, indicating that strange quarks were polarized oppositely to the up- and down-quarks and that they make a substantial contribution to the total, again in conflict with the simple quark model picture. Together with analyses of the $\pi N$ scattering that suggested a large contribution from strange quarks to the nucleon mass, these results suggested that strange quarks might also play a substantial role in other nucleon properties.

Kaplan and Manohar\cite{Kaplan:1988ku} subsequently pointed out that the use of WNC observables in lepton-nucleon scattering, in conjunction with information from purely electromagnetic (EM) scattering, could allow one to disentangle the individual $u$-, $d$-, and $s$-quark contributions to the nucleon electromagnetic structure. Shortly thereafter, Jaffe\cite{Jaffe:1989mj} observed that dispersive analyses of EM form factors suggested a considerably larger $\phi NN$ coupling than one would expect based on the OZI rule and that within the context of the vector meson dominance framework, one would then expect sizable strange quark contributions to the magnetic moment and charge radius of the nucleon.

On the experimental side, McKeown\cite{Mckeown:1989ir} showed that a \lq\lq strange magnetic moment" of the magnitude predicted by Jaffe could be observed in PV elastic electron-proton scattering, while Beck\cite{Beck:1989tg} noted that one could also probe the strange quark contributions to the nucleon electric form factors with PV electron scattering. The result of this activity was a nearly 20-year program of PV electron scattering experiments at MIT-Bates, Jefferson Lab, and Mainz. This effort built on the earlier work with $^{12}$C and $^9$Be as well as careful theoretical scrutiny of electroweak radiative corrections\cite{Musolf:1990ts,Zhu:2000gn}. In particular, it was noted that there exist sizeable hadronic effects in corrections to the nucleon axial vector amplitudes associated with so-called \lq\lq anapole moment" effects that do not enter the corresponding neutrino-nucleon interaction. Consequently, $\nu$-$N$ scattering provides the theoretically cleanest probe of axial vector strange quark effects.

The program of PV electron scattering ultimately showed that strange quarks play a relatively minor role in the nucleon electromagnetic structure, despite indications from a variety of hadronic effective approaches that had suggested otherwise. It now stands as a challenge to lattice QCD to reliably compute these small contributions, associated with pure sea quark degrees of freedom, and to explain the related impact strange quarks may have on nucleon spin and mass. At the same time, the PV program stimulated a new effort to measure the neutron distribution in lead\cite{Abrahamyan:2012gp}, exploiting the $\mathcal{O}(1)$ weak charge of the neutron compared to the suppressed proton weak charge (for a theoretical discussion, see, {\em e.g.}, \cite{Musolf:1993tb,Horowitz:1999fk} and references therein). In addition, the success of the PV technique lead to renewed interest in using PVES to test the SM weak neutral current interaction and search for indications of physics beyond the SM (BSM). Several experiments resulted, including a precise measurement of the PV asymmetry in M\o ller scattering at SLAC\cite{Anthony:2005pm} and elastic $ep$ scattering at Jefferson Lab~\cite{Armstrong:2012ps}. These measurements provide the most precise determinations of the scale-dependence of $\sin^2\theta_W$ that is predicted by the SM and are sensitive to possible BSM physics at the TeV scale. Looking to the future, more precise measurements of these asymmetries are planned for Jefferson Lab (M\"oller scattering) and Mainz (elastic $ep$), while lower-energy version of the original SLAC PV deep inelastic $eD$ experiment with broader kinematic coverage is also planned for Jefferson Lab. For reviews of the development and future prospects for PVES, see \cite{Musolf:1993tb,GonzalezJimenez:2011fq,Beck:2001dz,Beise:2004py}.

A parallel and interesting program of experiments and theoretical work has focused on parity-violation in purely hadronic reactions. Hadronic PV observables are sensitive to both CC and WNC interactions. Moreover, they are uniquely sensitive to the strangeness conserving component of the underlying quark-quark weak interaction, in contrast to the well-studied strangeness changing weak decays. As with the foregoing experiments, precise measurements of hadronic PV observables entail considerable challenges. In some nuclei, the effects of the PV interaction can be amplified by fortuitous aspects of nuclear structure, such as the presence of closely-separated opposite parity-states that lead enhanced parity admixtures in the nuclear wavefunctions. The theoretical interpretation of the measurements that have involved few-nucleon and many-body systems, is significantly more challenging than for the semileptonic or purely leptonic processes discussed above, due to the interplay of the weak interaction with the non-perturbative strong interaction. Because space limitations does not permit us to do justice to this interesting field here, we instead refer the reader to recent reviews on the topic, such as \cite{RamseyMusolf:2006dz}.

\section{Time-Reversal Tests}
\label{sec:trv}

\begin{figure}
\begin{center}
\includegraphics[width=0.4\textwidth]{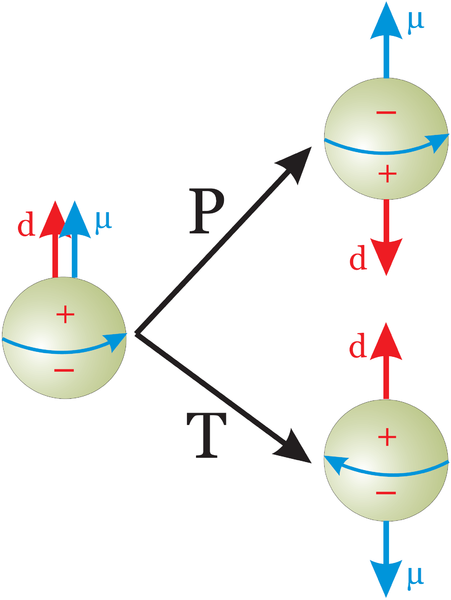}
\end{center}
\caption{Illustration of a system with permanent magnetic- $\mu$ and electric- $d$ dipole moments and the separate actions of parity (P) and time reversal (T).  In both cases, the relative orientation of the moments changes; thus both P and T would be violated if both $\mu$ and $d$ exist in a system. Figure courtesy A.~Knecht.} \label{fg:edm}
\end{figure}

As indicated earlier, the classic tests of combined P and T symmetry are searches for the permanent EDMs of the elementary particles and quantum bound states, such as the neutron or neutral atoms. A common semi-classical illustration is presented in Fig.~\ref{fg:edm}.  The magnetic moment, $\vec{\mu}$, defines an orientation, call it $\hat{z}$. If a permanent EDM $\vec{d}$ exists, it must be along or opposite $\hat{z}$.  The action of parity reverses the ``charges'' that create $\vec{d}$, but it does not affect $\vec{\mu}$.  The action of time reversal, reverses $\vec{\mu}$, but does not affect $\vec{d}$.  Therefore, an EDM is not invariant under P or T. Alternately, from a quantum mechanical point of view, the energy of a particle with spin ${\vec J}$ interacting with a magnetic and electric field is given by the Hamiltonian
\be
\mathcal{H} = - \frac{\mu}{J} {\vec J}\cdot{\vec B} - \frac{d}{J} {\vec J}\cdot{\vec E}\ \ \ .
\ee
The magnetic dipole term is even under P and T while the electric dipole interaction violates both of these symmetries individually.
Under the assumption of CPT invariance, the latter is not invariant under the product CP.

The experimental investigations generally exploit the original idea of Purcell and Ramsey\cite{Purcell:1950zz}, whose pioneering work launched the nearly six decade effort to observe an EDM. Interestingly, Ramsey and Purcell were originally motivated by the idea of testing P conservation in the strong interaction. Their experiment, launched before the $^{60}$Co and pion decay experiments, yielded a null result that was not published until after publications of the observation of PV in weak decays. In their 1957 paper, Ramsey, Purcell and Smith\cite{Smith:1957ht} noted that Lee and Yang had also mentioned that T-invariance would forbid an EDM, but their emphasis was primarily on testing P. Today, of course, EDM searches are motivated the new-physics implications of the implied CP-violation.

The SM expectations for non-vanishing EDMs arise from two sources. The electroweak CP-violation first observed in neutral kaon decays and later in the $B$-meson system implies a non-vanishing EDM of the neutron at the three-loop level in the presence of two strangeness-changing weak interactions involving the light quarks and gluons. For the electron, the corresponding prediction is further suppressed. The resulting predictions are several orders of magnitude below the present EDM limits as well as the expected sensitivity of the next generation of searches. The second source is possible CP-violation in the strong interaction associated with the \lq\lq $\theta$"-term in the QCD Lagrangian. Null results for neutron\cite{Baker:2006ts} and $^{199}$Hg EDM \cite{Griffith:2009zz}searches imply that the magnitude of the parameter ${\bar\theta}$ is no larger than $\sim 10^{-10}$, whereas one might na\"ively expect it to be of order unity. The resulting \lq\lq strong CP-problem" lead Peccei and Quinn\cite{Peccei:1977hh} to propose a new symmetry whose spontaneous breaking would imply the existence of an as-yet unobserved particle called the axion. Apart from its interest from a particle physics perspective, the axion is also a candidate for the cold dark matter that makes up roughly 23\% of the cosmic energy density. A number of axion searches have been carried out over the years, thus far yielding null results, but motivating new and more sensitive searches in the future.

In addition to SM CP-violation, EDMs provide powerful probes of BSM CP-violation. BSM scenarios typically predict new sources of CP-violation so the observation of such effects -- in addition to $\theta$-term CP-violation -- would constitute a significant discovery. In addition, explaining the excess of visible matter over anti-matter in the universe requires that there have existed BSM CP-violation sometime after the Big Bang and prior to the completion of electroweak symmetry-breaking when the universe was roughly 10 pico-seconds old. EDMs in particular probe the possibility that the abundance of visible matter was created during the latter era through \lq\lq electroweak baryogenesis" (for a recent review, see, Ref.~\cite{Morrissey:2012db}). EDM limits have already placed severe constraints on some electroweak baryogenesis scenarios, and the next generation is poised to test conclusively the most widely-considered supersymmetric versions. Thus, the observation of one or more EDMs could have profound implications for one of the outstanding problems lying at the interface of particle and nuclear physics with cosmology. At the same time, null results could point to alternative explanations that are less directly testable in laboratory experiments, such as baryogenesis via leptogenesis.

Despite dozens of efforts on a variety of atomic, molecular and particle systems, EDM tests use a common experimental theme. The system being tested is allowed to precess in a magnetic field with an aligned and anti-aligned electric field. The Larmor frequency is described by
\be
h\nu = 2(\mu_B B \pm d E).
\ee
A frequency shift that is proportional to the electric field strength and to the magnitude of an EDM would arise for the two relative field orientations: $\Delta h\nu = 4dE$.
It is basically that simple.  However, to reach the extraordinary precision of modern experiments, the alignment of electric and magnetic fields, the stability of the fields, and the reproducibility of the system under field rotations, along with a myriad of seemingly tiny issues, all enter.
The ``Ramsey separated oscillatory fields'' technique is usually employed in one form or another. It enables a very precise frequency-shift test.  Very briefly, we describe it for a neutron measurement.

A measuring volume is prepared with a magnetic field $\vec{B}$ along the $\hat{z}$ axis.  Neutrons are introduced into the cell with their spins polarized along $\hat{z}$. An oscillatory magnetic field, $B'$ is applied transverse to $\vec{B}$ at approximately the Larmor frequency.  In the rest frame of the neutron, it appears to tip the magnetic field over and the neutron spin spirals down by $90^\circ$ into the plane perpendicular to $\hat{z}$.  This ``$\pi/2$ pulse'' is removed and the neutron is allowed to precess freely for a fixed time interval. Meanwhile, the precision oscillator driving $B'$ continues to run.  Next, $B'$ is re-applied to the system for another $\pi/2$ pulse duration. If the Larmor and $B'$ frequencies are equal, the second $\pi/2$ pulse continues the action of the first pulse and the neutrons end up with their spins pointing down. If the frequencies are slightly different, the neutron spins have a probability of ending up or down,
determined by the phase difference accumulated between the Larmor and $B'$ oscillations during the time between the $\pi/2$ pulses. In practice, the $B'$ frequency
is chosen so that half of the neutron spins end up and half down; the number of neutrons in each spin state is counted separately. A frequency shift caused by an EDM
is then seen as a change in the relative number of spin up and down neutrons as the electric field is reversed.

The most sensitive absolute EDM limit is from the recent Seattle $^{199}$Hg EDM experiment, which obtained $|d(^{199}$Hg$)|~< 3.1 \times 10^{-29} e\cdot$cm$~(95\%$~(C.L.)~\cite{Griffith:2009zz} using a modified version of the above technique; they prepared and rotated the spin differently.  An EDM limit in this complex atomic system can be interpreted only after accounting for self shielding, the Schiff moment is measured, which nevertheless is very competitive in terms of nucleon-nucleon CP-violating interaction tests and new-physics limits.
In the neutron system, the ILL experiment presently holds the record, with
$|{d(n)}| < 2.9 \times 10^{-26} e\cdot$cm$~(90\%$~(C.L.)~\cite{Baker:2006ts}.  It is roughly equally  competitive with the Hg effort in terms of new physics reach.  Many next-generation efforts are being planned in the U.S. and abroad. The aim is typically up to a 100-fold improvement in sensitivity.

\section{Outlook}
\label{sec:out}
The study of P- and T-violation remains a vital area of research in nuclear physics and one that has significant implications for elementary particle physics, cosmology, and astrophysics. Studies of PV observables in weak decays, currently sensitive to ppt deviations from SM expectations, could reach another order of magnitude in sensitivity with the advent of the PERC neutron-decay detector in Heidelberg. The observation of a deviation at this level could be indicative of BSM physics, either entering through loops such as in supersymmetry (SUSY) \cite{Bauman:2012fx} or the exchange of heavy particles\cite{Cirigliano:2012ab}. As such, these studies could provide important information about the larger framework in which the SM resides, complementing what may be learned in the coming decade from the CERN Large Hadron Collider (LHC).

A future program of PV electron scattering experiments at Jefferson Lab and Mainz are poised for similar breakthroughs in sensitivity. For example, the PV M\o ller experiment proposed for Jefferson Lab would match the Z-pole sensitivity to the weak mixing angle, possibly providing a resolution to the present $\sim 3\sigma$ discrepancy between the values extracted from $A_\mathrm{FB}(b{\bar b})$ and LEP and the PV polarization asymmetries at SLAC. As with the weak decays, any deviations from the SM asymmetry predictions could point to SUSY (see \cite{RamseyMusolf:2006vr} and references therein), an additional neutral gauge boson ($Z^\prime$) \cite{Erler:2011iw,Li:2009xh,Chang:2009yw,Buckley:2012tc}, or some other BSM scenario and, if the corresponding particles are discovered at the LHC, help determine the underlying couplings to SM particles. At the same time, PV electron scattering will continue to provide a new window on poorly-understood aspects of nucleon structure, such as higher twist\cite{Mantry:2010ki,Belitsky:2011gz} and charge symmetry in parton distribution functions\cite{Hobbs:2008mm}.

The next generation of EDM searches involving the neutron, neutral atoms (such as Mercury, Radium, and Radon), and molecules (including YbF and ThO) -- and possibly searches involving the proton and light nuclei in storage rings -- are poised to improve on the present level of sensitivity by two or more orders of magnitude. If achieved, these experiments could probe BSM mass scales in the 10-50 TeV range, exceeding what can be accessed directly at the LHC, or provide evidence for a non-vanishing ${\bar\theta}$ parameter in QCD. Either way, the observation of a non-vanishing EDM would constitute a significant discovery, with the potential to provide new insights into the nature of what lies beyond the SM and to help to unlock the origin of matter.

In short, the study of P- and T-violation in nuclear physics continues to provide a unique window on the fundamental laws of nature, building on over five decades of significant experimental and theoretical advances. The history of this field of research is rich. Its future promises to be even more so.

\begin{acknowledgments}
This work was supported in part by DOE contracts DE-FG02-97ER41020 (DWH) and DE-FG02-08ER41531 (MJRM ), and by the Wisconsin Alumni Research Foundation (MJRM ).

\end{acknowledgments}

\end{document}